\documentclass{JHEP3}

\usepackage{amsmath,amsfonts}%
\usepackage{epsfig}

\newcommand{\bea}{\begin{eqnarray}}
\newcommand{\eea}{\end{eqnarray}}
\def\beann{\begin{eqnarray*}}
\def\eeann{\end{eqnarray*}}

\newcommand{\beq}{\begin{equation}}
\newcommand{\eeq}{\end{equation}}

\newcommand{\ba}{\begin{array}}
\newcommand{\ea}{\end{array}}

\def\ben{\begin{enumerate}}
\def\een{\end{enumerate}}

\def\4{\tilde }
\def\5{\bar }
\def\6{\partial }
\def\7{\hat }

\def\cL{{\cal L}}

\def\cN{{\cal N}}

\def\cP{{\cal P}}
\font\mybb=msbm10 at 12pt
\def\bb#1{\hbox{\mybb#1}}
\def\bR {\bb{R}}

\preprint{hep-th/0404239}
\title{Supertube domain-walls and elimination
of closed time-like curves in string theory}
\author{Nadav Drukker \\
The Niels Bohr Institute, Copenhagen University\\
Blegdamsvej 17, DK-2100 Copenhagen, Denmark.\\
\email{drukker@nbi.dk}}

\abstract{

We show that some novel physics of supertubes removes closed
time-like curves from many supersymmetric
spaces which naively suffer from this problem. The main claim is
that supertubes naturally form domain-walls, so while analytical
continuation of the metric would lead to closed time-like curves,
across the domain-wall the metric is non-differentiable, and the
closed time-like curves are eliminated. In the examples we study
the metric inside the domain-wall is always of the G\"odel type,
while outside the shell it looks like a localized rotating object,
often a rotating black hole. Thus this mechanism prevents the
appearance of closed time-like curves behind the horizons of
certain rotating black holes.
}


\begin{document}

\section{Introduction}

Many solutions of general relativity and supergravity exhibit
closed time-like curves, though they have sources that seemingly
are physical. One example is the G\"odel universe \cite{goedel},
which describes a rotating space with negative cosmological
constant. Some supersymmetric solutions of supergravity were found in 
\cite{gauntlett}, which are similar in many ways to the original metric 
of G\"odel, but without a cosmological constant, as  the metric of Som 
and Raychaudhuri \cite{somray}.

These constructions raised the question if string theory can live on 
spaces with closed time-like curves, and some interesting ideas related 
to holography were raised in \cite{horava}. Many other papers, including
\cite{harmark} -
\cite{Behrndt:2004pn} looked at these spaces, as well as other 
metrics with closed time-like curves in string theory and supergravity, and 
studied the dynamics of different objects in them. This is a new twist on 
the question if spaces that violate causality are good backgrounds for 
physics, see for example \cite{hawking,Herdeiro:2000ap,Herdeiro:2002ft}.

The metrics of the G\"odel type describe homogeneous spaces, which can 
be represented as rotating about any given point. Choosing a
preferred point there is a certain region around it which does not
contain closed time-like curves. This region is bounded 
by a surface made up of closed null curves, usually
called the ``velocity of light surface.'' The physics restricted only to this 
region is totally causal, and causality violation requires travelling
outside this domain.

Thus closed time-like curves may be prevented if the metric is
changed beyond the velocity of light surface, which can be done by putting
a domain-wall there. In a
previous paper \cite{we} it was shown that this happens naturally
in a certain example of a supersymmetric G\"odel universe with
rotation in a single plane\footnote{
This same idea was employed in general relativity for the original 
G\"odel metric in \cite{bonnor}.}. 
The purpose of this paper is to prove
that this mechanism is quite general, and in many examples
supertubes will form domain-walls and prevent the appearance of
closed time-like curves.

The G\"odel like universes are quite sick, as they have very
peculiar asymptotics. But there are many solutions of general
relativity and supergravity which are asymptotically flat, but
contain closed time-like curves. When examining such rotating
metrics it's quite common to find that the metric near the source
rotates fast enough to create closed time-like curves. Again one
may define a ``velocity of light surface'', outside of which causality is
protected. The simplest examples of such metrics are rotating
black holes, where the velocity of light surface is at the inner-horizon.

Applying our mechanism we will show that in certain examples a domain-wall 
of supertubes at the velocity of light surface will source these metrics. 
This will preserve the asymptotic form of the
metric, but will change it beyond the velocity of light surface. Across the
domain-wall the metric is continuous, but not differentiable, so
looking inside the domain-wall one would find a metric that has no
sources, is rotating, and whose analytical continuation would have
closed time-like curves beyond the domain-wall. Thus the metric
inside is just the G\"odel like spaces mentioned above.

Our main claim is thus that the regions with closed time-like
curves in G\"odel's space, as well in the rotating black holes are
the wrong analytical continuation of the metric. In fact one {\em
must} include a source that forms a domain-wall, thus the metric
will not be analytic but will continue from a piece of G\"odel
universe to a rotating asymptotically flat space.

The obvious question that would arise, is how can a supertube, 
which is a three-dimensional object, a tubular
D2-brane\footnote{ We will also look at other examples, which are
dual to the regular supertube, as well as the three charge one
constructed recently in \cite{Bena:2004wt}, which is a tubular
D6-brane. But none of those objects are nine-dimensional, as a
domain-wall should be.} fill out nine-dimensions and form 
a domain-wall? Usually in string theory D-branes are
localized objects, but we will show that a supertube becomes
delocalized as it approaches a velocity of light surface.

The statement that a D-brane is localized is based on the fact
that the scalars in the world-volume theory on the brane get a
VEV. The location of the D-brane, is just this VEV, and all
possible locations are described by the moduli space of the
relevant gauge theory. We will show that the entire velocity of light 
surface, though it has six dimensions transverse to the supertube is
actually a single point in the supertube moduli space. Thus a
supertube cannot sit at a given location on the velocity of light surface,
rather is smeared over it.

There is another example of such an effect in string theory,
dubbed the enhan\c{c}on mechanism \cite{enhancon}. There the naive
metric of the D6-brane wrapping a K3 surface has a naked
singularity, but by looking at the moduli space of the gauge
theory living on the brane one realizes that it actually cannot be
brought too close to the origin, and forms a spherical
domain-wall. Thus the naked singularity is replaced by the
domain-wall.

Our example is a very close analogy, with the D6-brane replaced by
the supertube (or one of its duals). In both examples we have the
same amount of supersymmetry, eight supercharges (or $\cN=4$ in
three dimensions), but while the gauge theory on the D6-brane is
well known, and the enhan\c{c}on locus is the $SU(2)$ point, the
physics on the supertube is a non-commutative gauge theory, with a
compact direction that becomes light-like (DLCQ) as the velocity of light 
surface is approached. As far as we know this field theory has never been
studied.

In the next section we will study the supertube as a probe of certain spaces 
with closed time-like curves. We will make some comments on the field
theory living on the brane, and calculate the metric on moduli
space. As we shall see the velocity of light surface will correspond to a
single point in moduli space, providing the justification for
smearing the supertube on that locus. In the following section we
apply this mechanism to several metrics, finding families of
metrics which have the G\"odel form behind the domain-wall and are
asymptotically flat rotating metrics outside. There are many such
metrics and we will not try to provide a complete survey of all of
them, instead illustrating the technique on certain interesting
examples. Perhaps the most interesting being the BMPV black hole
\cite{Breckenridge:1996is}. Finally we end with a discussion on
the relevance of this mechanism to causality protection in string
theory and to the physics of black holes.

\section{Supertube probe}
\label{sec:action}

\subsection{Supertubes}

Let us review the essential features of supertubes \cite{mateos}.
Those are bound states of D0-branes and fundamental strings with
angular momentum that take up a cylindrical shape. The most
convenient way to describe them is in terms of a cylindrical
D2-brane, which carries magnetic and electric fluxes corresponding
to the D0 and fundamental string charges respectively. Quite
remarkably, by adjusting the magnetic and electric field the
supertube can take any profile in the transverse space, and it
will still be supersymmetric. The supersymmetry projectors are
only those related to the D0-brane and fundamental string. The
system preserves therefore $1/4$ supersymmetry, or eight
supercharges.

In flat space the supertube is described by the Dirac-Born-Infeld 
action
\begin{equation}
\cL=-\tau_2 \sqrt{-\det(g+2\pi\alpha' F)}\,.
\end{equation}
$\tau_2=1/((2\pi)^2\alpha'^{3/2}g_s)$ is the D2-brane tension, 
$g_s$ is the string coupling, $g$ is the induced metric, and 
$F$ the field strength in the world-volume. As mentioned, we 
should turn on electric and magnetic fields, so taking the supertube 
extended in the $t$ and $y$ directions of space, and along a circle 
of radius $R$ with an angular coordinate $\varphi$, the ansatz is
\begin{equation}
2\pi\alpha'F=E\,dt\wedge dy+B\,dy\wedge d\varphi\,,
\end{equation}
the action for static configurations is
\begin{equation}
\cL=-\tau_2\sqrt{R^2(1-E^2)+B^2}\,.
\end{equation}
The integrals of the magnetic field and of the momentum conjugate 
to the electric field, $\Pi$, are conserved quantities
\begin{equation}
Q_s=2\pi\alpha'\int d\varphi\,\Pi
=2\pi\alpha'\int d\varphi\,\frac{\delta\cL}{\delta E}\,,
\qquad
Q_0=\frac{1}{2\pi\alpha'}\int d\varphi\, B\,.
\end{equation}
The energy per unit length of the tube is given by the integral of 
the Hamiltonian density
\begin{equation}
\tau=\tau_2
\int d\varphi\,R^{-1}\sqrt{(B^2+R^2)(\Pi^2/\tau_2^2+R^2)}
\geq\tau_0|Q_0|+\tau_s|Q_s|\,,
\end{equation}
where $\tau_0=1/(2\pi g_s\alpha'^{1/2})$ and 
$\tau_s=1/(2\pi\alpha')$ are the D0-brane and string tensions.
This bound is saturated for $\tau_2 R^2=Q_0Q_s$, and 
from that one can 
also derive the condition $E=1$. This condition is also necessary 
in order for the supertube to be supersymmetric.

The supertube carries angular momentum, which can be thought as 
the Pointing vector in the world-volume. Assuming the BPS bound 
is saturated, and all the D0-branes and string belong to the bound 
state, it is given by
\begin{equation}
J=Q_0Q_s=\tau_2R^2\,.
\end{equation}

In general we do not have to assume that there is a single D2-brane, 
but any number $N$. This means that from the D2-brane 
worldvolume point of view there is a $U(N)$ gauge theory, and the 
relations above will be modified. For a given radius the angular
momentum will be linear in the number of D2-branes and the total
charge $Q_0Q_s$ will be $N$ times larger still, because it is less
efficient to get angular momentum out of the non-Abelian fields
than from the Abelian ones. Thus we find
\begin{equation}
J=\frac{Q_0Q_s}{N}=\tau_2NR^2\,.
\label{bound}
\end{equation}
This issue is explained in detail in \cite{emparan}.

\subsection{Probe analysis}

As was shown in \cite{we}, supertube are interesting probes of 
certain type IIA supergravity solutions with closed time-like curves, 
specifically G\"odel universes. We wish to study this
probe further, and place it in some other geometries that have 
closed time-like curves, like some of the G\"odel universes
described in \cite{harmark}, or black hole solutions. 
Let us take the solutions of IIA supergravity of the 
supertube family given by \cite{paul,emparan}
\begin{equation}
  \begin{aligned}
 \label{supertube}
    ds^2=&-U^{-1}V^{-1/2}(dt-A)^2+U^{-1}V^{1/2}dy^2
          +V^{1/2}\sum_{i=1}^8 (dx^i)^2 \\
    B_2=&U^{-1}(dt-A)\wedge dy-dt\wedge dy\,, \\
    C_1=&-V^{-1}(dt-A)+dt\,, \\
    C_3=&-U^{-1}dt\wedge dy\wedge A\,, \\
    e^\Phi=&U^{-1/2}V^{3/4}\,.
  \end{aligned}
\end{equation}
Here $U$ and $V$ are harmonic functions in the eight $x^i$ 
directions and $A$ is a one-form satisfying the Maxwell equation. 
Later we will concentrate on specific examples corresponding to 
G\"odel universes with angular momentum in one or two planes, and to the 
two-charge black hole, but the probe analysis is actually much more 
general.

We consider a supertube in this background extended in the $y$ and $t$ 
directions and following some closed path in the transverse space. Let us 
label the periodic coordinate on the supertube by $\varphi$.
In the world volume we turn on electric and magnetic fluxes given by
\begin{equation}
2\pi\alpha' F=E\,dt\wedge dy+B\,dy\wedge d\varphi\,.
\end{equation}
In addition there are the components of the NS-NS $B$-field pulled back 
to the worldvolume
\begin{equation}
\cP[B_2]=\left(-1+U^{-1}\right)dt\wedge dy
+U^{-1}f\,dy\wedge d\varphi\,,
\end{equation}
where $f$ is the pull-back of the one form in the periodic direction 
$f\,d\varphi=\cP[A]$.

The metrics of the supertube type may have closed time-like curves 
when the space is rotating, so $A$ has an angular component. If the 
supertube is circular, and in a rotating plane, $f$ will be non-zero. The 
angular component of the metric is $V^{1/2}(r^2-f^2/(UV))$, and it 
can be negative. For example, the G\"odel metrics are given by 
constant $U$ and $V$ and $f=cr^2$, so when $r^2>UV/c^2$, this angular 
direction becomes time-like.

At low energies the dynamics of the supertube will be described by 
some gauge
theory. The simplest description of the system, as was done in 
\cite{bak} for a supertube in flat space, is in terms of
a non-commutative gauge theory, where one absorbs the effect of
the magnetic and electric fields in the open string metric and
non-commutativity parameters \cite{Seiberg:1999vs}. 
To that end first apply a gauge transformation that sets
the world-volume gauge fields to zero at the expense of the NS-NS
$B$-field. Effectively the total $B$ field induced on the world 
volume is (after setting $E=1$)
\begin{equation}
\hat B_2=\cP[B_2]+2\pi\alpha'F
=\left(U^{-1}\right)dt\wedge dy
+\left(U^{-1}f+B\right)\,dy\wedge d\varphi\,.
\end{equation}
This gives the open string metric and noncommutative parameter 
\cite{Kruczenski:2002mn}
(the lines and columns correspond to the $t$, $\varphi$ 
and $y$ directions)
\begin{equation}
 G^{ab}+\Theta^{ab}=
 \left(\frac{1}{g+\hat B_2}\right)^{ab}=
\frac{1}{B}
\begin{pmatrix}
-m&&-V^{1/2}&&f+Vr^2/B\cr
-V^{1/2}&&0&&1\cr
-(f+Vr^2/B)&&-1&&V^{1/2}r^2/B
\end{pmatrix}\,,
\label{open-string-metric}
\end{equation}
where
\begin{equation}
m=V^{1/2}\left(UB+2f+\frac{Vr^2}{B}\right)\,.
\label{m}
\end{equation}

With this we can immediately write the low energy action for the 
supertube, it's a non-commutative $U(1)$ gauge theory in three 
dimensions on a manifold with the open string metric 
(\ref{open-string-metric}). The bosonic part of the action will include 
the Yang-Mills part, and in addition seven scalars $X^i$ parameterizing the 
transverse fluctuations of the supertube with metric $\gamma_{ij}$
\begin{equation}
\cL=-\tau_2e^{-\Phi}\sqrt{-(g+\hat B_2)}
\left(\frac{1}{4}
G^{ab}G^{cd}F_{ac}F_{bd}
+\frac{1}{2}G^{ab}\gamma_{ij}D_aX^iD_bX^j\right)\,.
\label{action}
\end{equation}

\EPSFIGURE{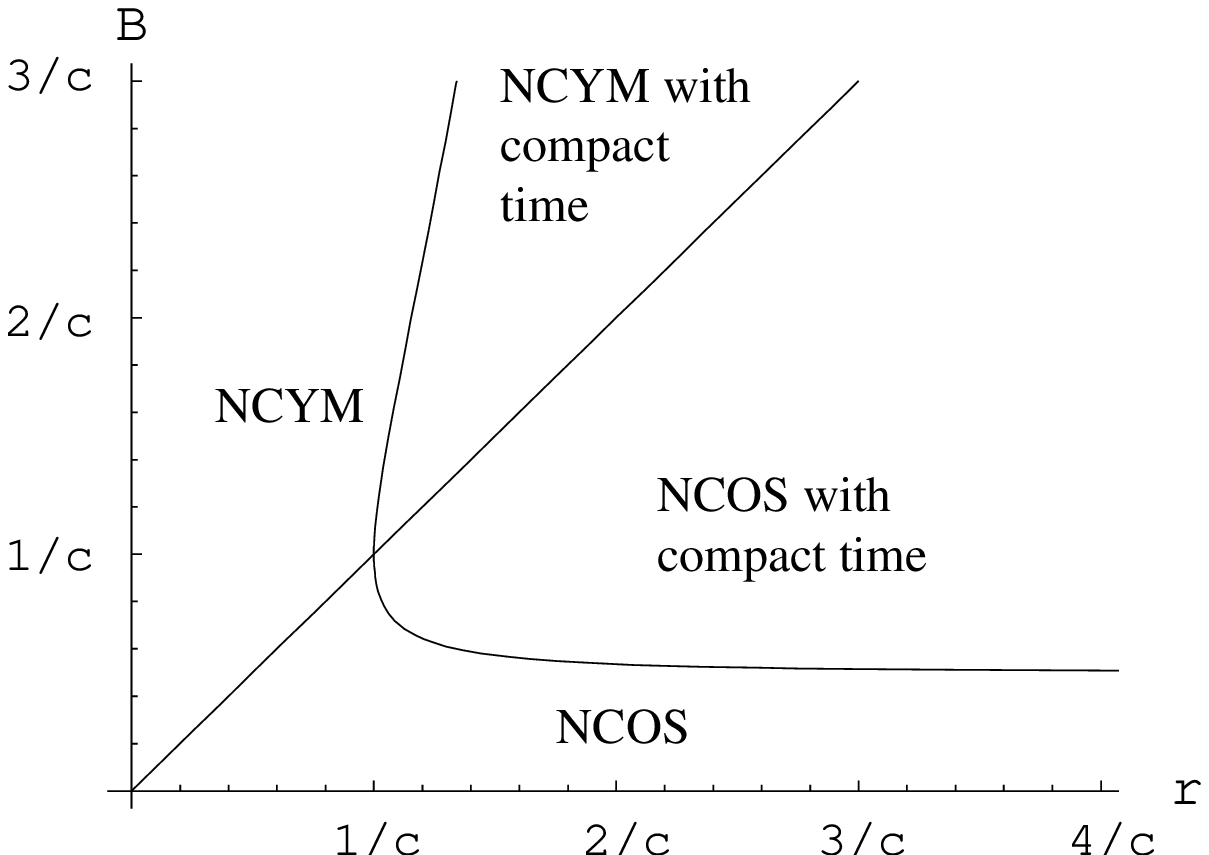, height=7cm}{The phases of the theory on the
supertube probing G\"odel's space.}

We can write the non-commutativity parameter as
$\Theta = B^2/(Vr^2)\,(-dt+(BU+f)d\varphi)\wedge dy$. The coordinate 
$y$ is always a space-like direction, the other direction is space-like for
$B^2>r^2V/U$ and time-like for $B^2<r^2U/V$.

In the open string metric (\ref{open-string-metric}) the square of 
the norm of the angular vector $\partial_\varphi$ is given by $m$, 
which we may rewrite as
\begin{equation}
G_{\varphi\varphi}=m
=\frac{Vg_{\varphi\varphi}}{B}
+\frac{V^{1/2}U}{B}\left(B+\frac{f}{U}\right)^2\,.
\end{equation}
This is the sum of the angular term in 
the closed string metric and a positive definite term. As long as the 
supertube wraps a space-like circle, $m$ is positive, but when 
the circle is a closed time-like curve, then for certain 
values of the magnetic field $m$ may be negative. This will 
correspond to a field theory on a manifold with a compact time.

Thus we have a two dimensional phase diagram with $r$ and $B$, and two
curves on it $B=(V/U)^{1/2}r$ (light-like non-commutativity) and $m=0$ 
(light-like compact direction).

Let us look at it in more detail in the case of G\"odel's universe, 
where $U=V=1$ and $f=-cr^2$ and is depicted in Figure 1.

\begin{itemize}
\item
For $B<1/(2c)$ as one varies $r$, the curve $m=0$ is not crossed, and 
there are two phases, for $r<B$ it's space-like noncommutativity 
and for $r>B$ time-like noncommutativity.

\item
For $1/(2c)<B<1/c$ at small $r$ there is space-like noncommutativity, 
then one crosses the $r=B$ curve to the time-like noncommutative 
phase, and finally at $m=0$ one crosses over to a phase with 
compact time-like direction.

\item
For $B>1/c$ one starts at small $r$ with space-like noncommutativity, 
then one reaches $m=0$, beyond which there is a compact time-like 
direction, and finally one reaches time-like noncommutativity.
\end{itemize}

We are quite accustomed to gauge theories with space-like
non-commutativity, and when the non-commutativity parameter is
time-like it's also possible to make sense of the theory by adding
extra open strings (NCOS theory) \cite{Gopakumar:2000na}. 
So it seems like as long as
$r<1/c$ all the probes are well behaved, with either space-like or
time-like non-commutativity. None of our probes develops any
sickness in that region, thus we think this region is safe and a
reasonable string background.

At larger radii there are certain probes that will cross the $m=0$
line, or have effectively a compact time-like direction. We do not
know how to make sense of those theories, at most we can have
theories with compact light-like directions (DLCQ, or discreet
light-cone quantization). The problem first occurs at $f^2=UVr^2$ for
the supertube with $B=-V^{1/2}U^{-1/2}r$, which has light-like 
non-commutativity
in this compact direction. This case is particularly simple,
because all the non-commutativity is in the $(\varphi,y)$ plane.

One approach to understand what happens to the supertube as it
nears the velocity of light surface, or the point of DLCQ is to study the
DLCQ limit of non-commutative gauge theories. In regular field
theories this limit is a very subtle one, because an infinite
number of zero-modes appear. This issue is much less 
severe for non-commutative field theories.

Let us recall the problem \cite{simeon}, if the periodic direction $\varphi$
becomes light-like, any field configuration that depends only on $\varphi$ will 
be a zero mode. It's just a free wave, and it can have arbitrary $\varphi$
dependence, it is then hard to control the perturbative expansion
with all those zero modes running around loops. If this compact
direction is non-commutative we may use the operator formalism,
then instead of our fields being functions of this direction, they
are operators on a Hilbert space where $\varphi$ and another
direction (in our case $y$) are conjugate variables. This means
that the $y$ dependence and $\varphi$ dependence of our field are
correlated, so there are no longer the zero-modes that are
functions only of $\varphi$, but do not depend on $y$.

One would therefore expect there to be at most a single zero mode for 
every field, and not an infinite degeneracy of them, and the perturbative
expansion will be well defined.

For practical purposes, though, the operator formalism is not very
convenient, and it is much easier to take the fields to be regular
functions, but add the necessary phase in the Feynman diagrams to
represent the Moyal product. This way the zero modes will show up
again, but they will still not lead to divergent graphs. In
non-commutative gauge theory the phases are always odd function of
the momentum, typically $\sin(p_\mu\Theta^{\mu\nu}k_\nu/2)$
\cite{Hayakawa:1999zf,Matusis:2000jf}. Since
the zero modes do not carry momentum in one of the non-commutative
directions, this phase factor kills the terms that would otherwise
be divergent.

This very peculiar type of theory, the DLCQ limit of
non-commutative field theory, should represent very interesting
physics, but we will not pursue it here.\footnote{the DLCQ of some six dimensional non-commutative 
theories were studied in \cite{Aharony:1997an,Aharony:2000gz}.}
Instead we will study the
moduli space of supertubes near this point where the compact
direction becomes light-like. This will turn out to be sufficient
for out purposes.

\subsection{Moduli space}

Our main claim is that when supertubes approach the velocity of light 
surface, they delocalize, forming a domain-wall on the velocity of light 
surface. This is very similar to the enhan\c{c}on mechanism, where a
D6-brane wrapping a K3 surface becomes delocalized over a
2-sphere, again forming a domain-wall \cite{enhancon}. In that
example the gauge theory was quite well understood, and the
enhan\c{c}on point corresponds to a point of enlarged gauge
symmetry.

The problem of a non-commutative gauge theory in the limit as the
compact direction becomes light-like is not as well understood, and
as stated above, will not be studied here in detail. But this will
not be required for what we are trying to prove, it will be enough
to look at the moduli space near this point of compact time.

Recall that the position of a D-brane is described in the
world-volume theory by the vacuum expectation value of some scalar
fields. It is usually the case that D-branes are localized,
and cannot be smeared (if their dimensionality is not too small),
and this corresponds to the fact that the scalar fields should
always get a VEV. But this in fact breaks down in the case of the
enhan\c{c}on, where the scalar fields are W-bosons, and clearly do
not get VEVs anymore in the enhanced symmetry point.

This can be seen by studying the moduli space of the theory on the
D-brane. One may deform the shape of the supertube without breaking 
supersymmetry \cite{Bak:2001xx,Mateos:2001pi,Bak:2002wy}, but 
such deformations might change the total charges 
carried by the supertube, those of D0-branes, fundamental strings and 
angular momentum. It is still possible to consider those deformations, assuming 
the supertube is in a bath of other supertubes, D0-branes and strings that 
can absorb the charge difference, but we will try to avoid this, and focus 
only on real moduli.

If the supertube carries the maximal angular momentum for given charges, 
its shape cannot be deformed. In the case of a single angular momentum 
it will have a circular geometry (also in curved space it will be a circle in 
the coordinates $x^i$ we are using). Then the only moduli are rigid translations 
of the supertube, which we will focus on.

The distance between two D-branes in space-time is given
by the difference between the scalar VEVs, but the distance in
field space is given by the metric on moduli space. 
This metric can be readily extracted from the low
energy effective action (\ref{action}), and is just proportional
to the kinetic term for the scalars
\begin{equation}
ds^2=M_{ij}dX^i dX^j\,.
\label{modmet}
\end{equation}
Here $X^i$ represent the center of mass of the tube and the coefficient $M$ 
is the average value of $m$ (\ref{m})
\begin{equation}
M_{ij}=\frac{1}{2\pi}\int d\varphi\, 
\frac{m}{B}e^{-\Phi}\gamma_{ij}\sqrt{-(g+\hat B_2)}\,.
\end{equation}

As a first example let us look at some G\"odel-like spaces. In that case 
$U=V=1$ and $A=-c_{ij} x^i\, dx^j$ with a skew-symmetric 
matrix $c_{ij}$. We take the supertube to lie in the $(7,\,8)$ plane, 
centered at $(X^7,\,X^8)$, and with a radius $R$, or explicitly we use 
the embedding
\begin{equation}
x^7=X^7+R\cos\varphi\,,
\qquad
x^8=X^8+R\sin\varphi\,.
\label{embed}
\end{equation}
The pullback of $A$ to the world-volume is then
\begin{equation}
f=-c_{78}(R^2+RX^7\cos\varphi+RX^8\sin\varphi)\,.
\end{equation}
This gives the kinetic term for the zero modes
\begin{equation}
M_{ij}=\frac{1}{2\pi}\int d\varphi
e^{-\Phi}\delta_{ij}\left(B+\frac{R^2}{B}+2f\right)
=e^{-\Phi}\left(B+\frac{R^2}{B}-2c_{78}R^2\right)\delta_{ij}\,.
\end{equation}

Thus the value of $M$ is independent of the position of the supertube, which 
is just a manifestation of the homogeneity of those G\"odel metrics. It seems 
like no interesting phenomena can emerge here, but still, as discussed above, 
$M$ may vanish or turn negative for certain values of $c_{78}$, $R$ and $B$. 
For all positive values of $M$ the moduli space is just a copy of $\bR^8$, 
like in flat space \cite{Bak:2004rj}, but 
when $M=0$ the metric on moduli space vanishes, so the moduli space 
shrinks to a single point.

If we allow the supertube to absorb charges, that would change $R$ and $B$
and we would find a much larger moduli space. A slice through this space 
with circular tubes of fixed values of the charges would be a copy of 
$\bR^8$, except for a singularity where $M=0$.

But since we are not assuming that $R$ and $B$ can be modified dynamically, 
those are regarded as parameters, not moduli. Still we can look at supertubes 
with different values of these parameters, and compare their dynamics. 
$M$ should be considered the inertial mass of the supertube, the smaller it is the 
easier it is to move the supertube around. When it vanishes, the supertube 
has no inertial mass and can no longer be localized. If the moduli space is 
a single point, one cannot consider supertubes at different positions, and 
the only possibility is that the tube is in a superposition all over space.

This establishes our main
claim that there is only a single theory for a supertube on the
velocity of light surface, and therefore the supertube is delocalized and
smeared over it. 
This happens for a tube of radius $R=1/c_{78}$, while for larger radii $M$ 
will be negative, which is unphysical. We conclude that a consistent 
background will have a domain-wall at this value of $R$ built of such 
smeared supertubes.\footnote{If one expands the action beyond the 
quadratic level there may be terms of higher order in $m$, which could 
in principle change the dynamics. The cases we will concentrate on below 
have $m=0$ for all $\varphi$, so all the higher order terms will also 
vanish and will not remove the singularity in moduli space.}

\subsection{Other backgrounds}

Above we studied what happens to a supertube as it approaches the
velocity of light surface in some metrics of the G\"odel type, but 
the same analysis is valid for many other backgrounds
with closed time-like curves. We saw that the low energy dynamics of the
supertube are described by a non-commutative supersymmetric three 
dimensional gauge theory. The scalars in the gauge theory 
parameterized a space that depended on the background.

In other backgrounds a lot of the same will still apply, the supertube 
will be described by a non-commutative gauge theory with eight 
supercharges, and as it approaches the velocity of light surface 
the compact direction becomes light-like. The metric on moduli space 
will be different, but still will be proportional to $M$, as before. Thus we
will arrive at the same conclusion that the supertube will
delocalize on the velocity of light surface, forming a domain-wall of the 
appropriate geometry.

We will study supertubes in other geometries below. As opposed to 
the G\"odel example, those metrics will not be homogeneous, and 
the moduli space will not be flat. Instead $M$ will vary and vanish only 
at some singular points. So everywhere in space supertubes can be 
localized, except when we reach the singular point, and they have to 
delocalize into a domain-wall.

\subsection{Validity of probe approximation}

It is important to establish the range of validity of the probe
analysis. The main requirement is that the back-reaction of the
supertube probe is small compared to the background. In all the
discussion above we assumed that the supertube is a single
D2-brane, carrying some D0 and fundamental string charges. If the
velocity of light surface we are studying is large, the supertube will
carry a macroscopic amount of charge, and therefore could
back-react on the geometry.

As was shown in the case of the three dimensional G\"odel space in
\cite{we}, and will be shown in other examples below, the metric
one gets from the back-reaction inside a shell of supertubes is one
of those G\"odel metrics. So we should really think of the probe
calculation not inside a homogeneous space, instead inside one of
those domain-wall solutions. For the probe approximation to be
valid we have to assume that the domain-wall is much heavier than
the probe, and the way to achieve that is by assuming the shell is
made up of a large number of supertubes, while it's probed only by
a single one.

So we have shown rigorously that if we have a shell made up of a
large number of supertubes, as an extra supertube approaches it,
it will dissolve into it, spreading homogeneously over the entire
domain-wall. Thus we wish to claim that the domain-wall is not an
approximate notion from a distribution of a large number of
supertubes. Each of the constituent supertubes is spread over the
entire velocity of light surface forming a continuous object.

While we can make a rigorous statement only for a large number of
supertubes, the same may be true for arbitrarily small numbers of
supertubes, in particular one. So a single supertube smeared into
a domain-wall may be a legitimate string theory background. While
we cannot substantiate this claim, the analysis above seems to
make it self consistent.

\section{Domain wall metrics}
\label{sec:metrics}

\subsection{Generalities}

We will apply our procedure to a few examples of known spaces with
closed time-like curves by adding domain-walls made out of smeared
supertubes. In the examples we consider the domain-walls will have
the topology $S^{k-1}\times \bR^{10-k}$, with $k=2,\,4$.

The sources are invariant under translations in $\bR^{10-k}$ and
under rotations in $\bR^k$ (if we ignore the angular momentum that
breaks the rotational symmetry) . Like in electro-magnetism the
solution outside the source will not depend on the radius of the
source, so the metric will look like that of a rotating point
source in $\bR^k$. Natural candidates are rotating black holes.
Those spaces usually have closed time-like curves, and a velocity of light 
surface with the topology $S^{k-1}\times \bR^{10-k}$. This is where
we will place the supertube source.

Inside the supertube there will be another metric, with the same
symmetries. Since the metric should be continuous across the
domain-wall, the interior metric will also have a velocity of light surface
exactly at the same location. So both the inside and outside
metrics agree on the location of the domain-wall. The
inside metric describes a space that by naive analytical
continuation would have had closed time-like curves outside a
certain radius. Natural examples of such metric are spaces of the
G\"odel type. If there was no rotation the metric inside the
domain-wall would have been flat space (as in the case of the
enhan\c{c}on), but the angular momentum carried by the supertubes
can be seen inside (like the magnetic field inside a solenoid). So
the metric inside should be close to a rotating flat space, and
G\"odel fits this description quite well.

All our simple examples will follow this pattern, with an interior metric
of the G\"odel type, a domain-wall, and outside a rotating black hole
metric.

One can consider more complicated examples, with concentric supertube
shells, where the inner most region will look like G\"odel, the outside
will be a black hole, and the extra regions in the middle will look like
black holes in G\"odel spaces.

It is possible to consider other topologies for the domain-wall.
One example is $S^1\times S^2\times\bR^6$, which may be
relevant for black ring metrics 
\cite{Emparan:2001wn,Elvang:2003yy,Elvang:2003mj}.

\subsection{Metrics with one angular momentum}

In \cite{we} it was shown that the three dimensional G\"odel solution
of supergravity can be regarded as the interior of a supertube domain-wall.
Let us briefly recall the construction. 

To form a piece of three dimensional
G\"odel universe we have to take a source that forms a domain-wall
of geometry $S^1\times\bR^8$. So we use polar coordinates $r$ and
$\phi$ in the $(x^7,\,x^8)$ plane.
With this source the metric for $r>R$ is described by (\ref{supertube}) with
\begin{equation}
    U=1+Q_s\ln\frac{r}{R}\,,\qquad
    V=1+Q_0\ln\frac{r}{R}\,,\qquad
    A=-R\,d\phi\,.
 \label{G3metout}
\end{equation}
$Q_s$ and $Q_0$ are respectively the fundamental string and
D0-brane charge densities. The angular momentum was set so at the domain 
wall, at $r=R$, the angular direction is null.

Inside the domain-wall, for $r<R$, this is continuous to
\begin{equation}
    U=1\,,\qquad
    V=1\,,\qquad
    A=-\frac{r^2}{R}d\phi\,.
 \label{G3metin}
\end{equation}
which gives a metric that looks like the three dimensional G\"odel universe
\begin{equation}
  \begin{gathered}
    ds^2 = -\left(dt+\frac{r^2}{R}d\phi\right)^2+dr^2+r^2d\phi^2
      +dy^2 +\sum_{i=4}^9 (dx^i)^2 \\
    H_3=2cr\,dr\wedge d\phi\wedge dy\,,\qquad
    F_2=-2cr\,dr\wedge d\phi\,,\qquad
    F_4=2cr\,dt\wedge dr\wedge d\phi\wedge dy\,.
  \end{gathered}
 \label{G3metric}
\end{equation}

Since the full metric does not have closed time-like curves we are 
guaranteed that no supertube will suffer from the problem discussed before, 
of having a compact time direction in the three-dimensional gauge
theory. At most we can find a compact time-like direction if the 
supertube coincides with the domain wall and has $B=R$. So let us consider 
the moduli space of a supertube with this radius and magnetic field, it 
can be translated in the transverse six directions as well as the plane it's 
in. We use the same embedding as before (\ref{embed})
\begin{equation}
x^7=X^7+R\cos\varphi\,,
\qquad
x^8=X^8+R\sin\varphi\,.
\end{equation}
The metric on moduli space is given by
\begin{equation}
ds^2=M\left[(X^7)^2+(X^8)^2\right]\sum_{i=1}^8(dX^i)^2\,.
\end{equation}
The coefficient $M$ is a function of the distance of the supertube from 
the domain-wall, and is given by the integral
\begin{equation}
M=\frac{1}{2\pi}\int d\varphi
\frac{BV}{U}\left(UR+V\frac{r^2}{R}+2f\right)\,.
\end{equation}
When the supertube is far away from the origin it will not cross the domain 
wall and to perform the integral one uses the external metric and the 
relations $r=(x^7)^2+(x^8)^2$ and 
$f=-(R+X^7\cos\varphi+X^8\sin\varphi)R^2/r^2$.
The resulting expression is quite complicated, and not illuminating.

It is more interesting what happens to the supertube as it nears the origin. 
To simplify the expressions let us set $X^8=0$, then if 
$X^7\ll R$ we may expand outside the domain-wall
\begin{equation}
 U\sim1+Q_s\frac{X^7}{R}\cos\varphi\,,
 \qquad
 V\sim1+Q_0\frac{X^7}{R}\cos\varphi\,,
 \qquad
 f\sim-R+X^7\cos\varphi\,.
\end{equation}
Recall that the same supertube inside a homogeneous G\"odel had $M=0$, 
so we only have to integrate the difference between the above and the 
naive continuation of the 
G\"odel metric outside the wall. At the linear approximation half the tube 
is inside and half outside. So we get
\begin{equation}
M\sim\frac{R}{2\pi}\int_{-\pi/2}^{\pi/2}d\varphi
\left(Q_s+Q_0+4\right)X^7\cos\varphi
=\frac{R}{\pi}\left(Q_s+Q_0+4\right)X^7\,.
\end{equation}
As expected it vanishes at the origin, so the supertube becomes smeared in 
the transverse six dimensions, implying the construction was consistent. 
Instead of studying this system further we turn now to the case with two 
angular momenta, which is much more interesting.

\subsection{Metrics with two angular momenta}

The three dimensional example was already studied in \cite{we},
and it suffers from the fact that the metric outside the domain-wall 
is not asymptotically flat. Let us therefore shift to another example, 
with rotation in two planes. There is a simple geometry that is
asymptotically flat, carries D0-brane and fundamental string charges and
angular momentum in two planes. It is dual to the 2-charge rotating black
hole made out of D1 and D5-branes, and is given by the general
ansatz (\ref{supertube}) with
\begin{equation}
    U=1+\frac{Q_s}{r^2}\,,\qquad
    V=1+\frac{Q_0}{r^2}\,,\qquad
    A=-\frac{J}{2r^2}\sigma_L^3\,.
 \label{5metout}
\end{equation}
$Q_s$ and $Q_0$ are the fundamental string and D0-brane charge
densities, as can easily be verified from Gauss's law. Since the
metric in nontrivial in five space-time directions, $r$ is the
radial coordinate in $\bR^4$, and in the gauge field $\sigma^3_L$ 
is a left-invariant one-form on $S^3$. If we use the polar coordinates 
$(\eta,\,\phi)$ in the $(X^5\,,X^6)$ plane and
$(\zeta,\,\psi)$ in the $(X^7\,,X^8)$ plane, so $r^2=\eta^2+\zeta^2$, 
we can write $\sigma^3_L=2\eta^2/r^2\,d\phi+2\zeta^2/r^2\,d\psi$.

The solution above carries angular momentum in the two planes and from 
the asymptotic form of the metric we can read off its value $J_L=-J$.

Far away from the origin this space is perfectly causal, but close
to the origin there are closed time-like curves. The purely angular
part of the metric is given by
\begin{equation}
 -U^{-1}V^{-1/2}A^2
 +V^{1/2}\left(d\eta^2+\eta^2\,d\phi^2
 +d\zeta^2+\zeta^2\,d\psi^2\right)\,,
\end{equation}
and at the radius $R$ where $UV=J^2/R^6$ it is proportional to
$d\eta^2+d\zeta^2+\eta^2\zeta^2/r^2(d\phi-d\psi)^2$. This metric 
is degenerate, so there are null curves on the $S^3$ at this radius, given 
by the direction $\partial_\phi+\partial_\psi$. This direction is 
the fibre of the Hopf fibration, so at the critical radius the $S^3$ degenerates 
to an $S^2$, while at smaller radii this direction becomes time-like.

We found therefore that this geometry has a velocity of light surface of
geometry $S^3\times\bR^6$. According to our general prescription
we should not trust the metric beyond this radius, instead we
should put the sources on this sphere. One may think of the source as 
supertubes that follow the null curves on this domain-wall, or the Hopf 
fibres. Thus the domain-wall is 
constructed from supertubes with geometry $S^1\times\bR^2$ fibred
over $S^2\times\bR^4$.

There are a few tests to check this construction. First compute what the 
interior metric is. Then see that this domain-wall does indeed give the
above desired geometry. Finally we will probe this metric with different 
supertubes and see that they will delocalize as stated.

Since there are no more sources in the interior it should be a
vacuum solution that is continuous to the one above across the
domain-wall. Such a solution exists, and is given by the ansatz
(\ref{supertube}) with
\begin{equation}
    U=1+\frac{Q_s}{R^2}\,,\qquad
    V=1+\frac{Q_0}{R^2}\,,\qquad
    A=-\frac{Jr^2}{2R^4}\sigma_L^3\,.
 \label{5metin}
\end{equation}
This is a deformation of the five dimensional G\"odel universe of
type IIA (eq (2.46) of \cite{harmark}). The only difference being
that $U$ and $V$ are not unity, but other constants.

The general supertube ansatz is described by harmonic functions
and a Maxwell field, so all the solutions have simple analogs in
electromagnetism. The three dimensional example in the previous
section is analogous to a charged solenoid. The harmonic functions
outside the domain-wall are the same as one would take for a
charged cylinder in four dimensions. The gauge field gives a
constant magnetic flux inside. The example at hand also has an
analog, in electromagnetism in five dimensions.

Since there are nowhere vanishing vector fields on $S^3$, one may
put an electric current on it. We will choose the current to run
along the fibres of the Hopf fibration, and the current
density will be proportional to the dual one-form $\sigma_L^3$.
For our analogy the shell should also carry a uniform charge
distribution. The solution of this five-dimensional
electromagnetic problem is given by the expressions
(\ref{5metout}) and (\ref{5metin}). Inside the domain-wall the
scalar potentials are constant, but there is a magnetic field, and
outside there are scalar potentials, as well as a decaying
electromagnetic field.

We should take some care to normalize the charges of this
solution. The location of the domain-wall is at the velocity of light 
surface at a radius $r=R$, and if we assume that $Q_s,Q_0\gg R^2$
this gives the condition $J^2=R^2Q_sQ_0$. Recall (\ref{bound})
that if the branes make up $N$ bound states (i.e. $N$ supertubes),
the bound on the angular momentum is $J\leq Q_0Q_s/N$, or $J\leq
R^2N$. We see that our solution saturates this bound.

We may look at a more general solution, where we put the domain-wall 
at a larger radius. Let us still take the same form for the
exterior metric, and by continuity find the same form for the
metric inside. The only difference is that the radius $R$ will be
greater than before. Note that in the interior metric the rotation
is proportional to $A=-Jr^2/(2R^4)$, and due to the negative
power of $R$, this G\"odel space in the interior has a smaller
rotation parameter than the case above. This solution will be
totally causal, with no closed time-like curves either inside or
outside of the shell. But now there are also no closed null
curves, and the supertubes are no longer at a velocity of light surface, so
the bound above is not saturated, instead $J^2<R^2Q_sQ_0$. This
means that not all the constituent D0-branes or fundamental
strings are part of the bound state. Also we cannot justify this
construction by the analog of the enhan\c{c}on mechanism, and one
should think of the supertubes as localized objects and are smeared only 
if one takes a continuum limit.

Given the general ansatz (\ref{supertube}), it's quite obvious that the 
metric we have written down is a solution, still it's instructive to check 
the junction conditions across the domain-wall \cite{israel} to verify that 
we have the correct source. In the case of the enhan\c{c}on this was 
done in \cite{myers}, and we follow their calculation. One first 
calculates the jump in the extrinsic curvature, given by
\begin{equation}
\gamma_{\mu\nu}=\frac{1}{2\sqrt{\tilde g_{rr}}}\left(
\partial_r \tilde g_{\mu\nu}^-
-\partial_r \tilde g_{\mu\nu}^+\right)\,.
\end{equation}
Here $\tilde g^\pm$ is the metric in the Einstein frame outside (+) and 
inside (-) the shell. The stress-energy tensor across the junction is simply
\begin{equation}
S_{\mu\nu}=\frac{1}{\kappa^2}
\left(\gamma_{\mu\nu}-\tilde g_{\mu\nu}\gamma^\rho_\rho\right)\,.
\end{equation}
The coupling in front, $2\kappa^2=(2\pi)^7\alpha'^4g_s^2$, is 
the Newton constant in ten dimensions.
Note that the tensors $\gamma$ and $S$ are defined on the domain-wall, so 
the indices $\mu\,, \nu\,, \rho$ take only nine values (all the directions 
excluding $r$).

The intermediate steps in the calculation are rather messy, but the final result 
is simple in particular if we choose to write the stress-energy tensor with 
upper indices. The only non-zero components are then
\begin{equation}
\begin{aligned}
S^{tt}=&\frac{Q_s+Q_0+2(R^2Q_sQ_0-J^2)/R^4}
{\kappa^2 g_{rr}^{3/2}R^3}\,,
\\
S^{t3}=&\frac{2J/R^2}{\kappa^2 g_{rr}^{3/2}R^3}\,,
\\
S^{yy}=&\frac{Q_s}{\kappa^2 g_{rr}^{3/2}R^3}\,.
\end{aligned}
\end{equation}
The last line is proportional to the fundamental string density, which 
is smeared over the three-sphere whose volume is 
$2\pi^2(R\sqrt{g_{rr}})^3$ (and we are using the normalization where 
charges are divided by the volume of a unit sphere). 
The second line corresponds to the angular momentum carried by the 
supertube, and the index $3$ corresponds to the direction of $\sigma_L^3$, 
where the angular momentum is, but there is no component in the 
$S^{33}$, or any of the other sphere directions. Finally 
the first line corresponds to the total energy density, which includes the 
contribution of the fundamental strings ($Q_s$) and D0-branes ($Q_0$). 
The other term vanishes when the supertube source carries maximal angular 
momentum $J^2=R^2Q_0Q_s$. To understand it note that the $G^{tt}$ 
component of the open string metric (which is proportional to $m$) 
represents the extra mass a supertube 
had over the tension of its constituents. We wrote it as the sum of two 
terms, one a positive, and another one,  $UVR^2-J^2$. As 
the first term vanishes for supersymmetic configuration, 
we are left with the second one, which 
to leading order is the same as the access energy in the calculation above.

Let us now consider a supertube in this domain-wall geometry. First taking 
a planar tube in the $(X^5,\,X^6)$ plane, and we consider the 
most interesting case where the radius coincides with the size of the domain 
wall, and set the magnetic field to $B^2=R^2(Q_0+R^2)/(Q_s+R^2)$, which 
is the value where the open string metric becomes degenerate at the 
velocity of light surface. As before we will not 
write down the full expressions, just concentrate on the behavior near the 
domain-wall. In this case we may displace the supertube in the plane, but 
also in the $X^7$ and $X^8$ directions. Labeling $\eta$ the displacement 
of the center of the supertube in the plane, and $\zeta$ the displacement 
in the other direction a simple calculation yields
\begin{equation}
m\sim
\begin{cases}
{\displaystyle 
\sqrt{V}
\left(\frac{R^2}{B}-\frac{Q_sB}{R^2}+\frac{2J}{R^2}\right)
\left(2\frac{\eta}{R}\cos\varphi+\frac{\zeta^2}{R^2}\right)
-\frac{J\eta}{R^3}\cos\varphi\,,}
&{\displaystyle 
-\frac{\pi}{2}<\varphi<\frac{\pi}{2}\hbox{ (outside)}}\\\\
{\displaystyle 
-\frac{J\eta}{R^3}\cos\varphi\,,}
&{\displaystyle 
\frac{\pi}{2}<\varphi<\frac{3\pi}{2}\hbox{ (inside)}}
\end{cases}
\end{equation}
To get this we used the relation $UVR^6=J^2$ at the domain wall. Now we 
can do the integration over $\varphi$ and replace for $B$, which gives
\begin{equation}
 M\sim\frac{Q_0+R^2}{(Q_s+R^2)^2}
 \left(Q_sQ_0+2R^2(Q_0+Q_s)+3R^4\right)
 \left(\frac{4\eta}{R}+\frac{\pi\zeta^2}{R^2}\right)\,.
\end{equation}
If we took the near-horizon limit, where $U=Q_s/r^2$ and $V=Q_0/r^2$, 
the expression would simplify to
\begin{equation}
M\sim \frac{Q_0^2}{Q_s}
\left(\frac{4\eta}{R}+\frac{\pi\zeta^2}{R^2}\right)\,.
\end{equation}

In both cases we find the expected result, that the metric on moduli 
space degenerates as the supertube reaches the velocity of light surface
($\eta=\zeta=0$). So if we bring in a 
supertube from infinity, it will naturally dissolve into the domain wall, and 
smear in the transverse $\bR^4$ direction. This is not enough to justify the 
domain wall, where we assumed that the supertubes are also smeared into 
an $S^3$, the reason being that our supertube carried angular momentum 
only in the $(X^5,\,X^6)$ plane. To rotate it into the other plane, we 
have to enlarge the moduli space to allow a change in the angular momentum 
by coupling to some external object. Otherwise we can consider more 
general supertubes that carry angular momentum in two planes.

One possibility is to consider the system comprised of a pair of supertubes, 
one in the $(X^5,\,X^6)$ plane and the other in the $(X^7,\,X^8)$ plane. 
The combined system carries both angular momenta, and for appropriate 
charges will have vanishing $M$ at the velocity of light surface. If the angular 
momenta in the two planes are equal, conservation of charge will not prevent 
us from rotating the two planes into each other and the combined system 
can be smeared into the $S^3$. But this rotation cannot 
take place smoothly as the two supertubes, which are disconnected, 
will have to exchange angular momentum.

Another possibility is to consider some objects constructed in 
\cite{Mateos:2002yf}. The idea is for 
the supertube to follow a curve that winds circles in the two planes. One 
takes the ansatz
\begin{equation}
X^5+iX^6=\eta e^{in\varphi}\,,
\qquad
X^7+iX^8=\zeta e^{im\varphi}\,.
\end{equation}
If $n=m$ this will be a circle in another plane, and carry only one 
independent angular momentum. But as long as $n\neq m$, this configuration 
carries two angular momenta. It has $J_1=n\eta^2$ in the $(X^5,\,X^6)$ 
plane, and $J_2=m\zeta^2$ in the second plane.

This tube wraps a curve  on a two-torus inside a three-sphere, and there is 
a zero mode for rotating it around the torus by the modification 
$X^5+iX^6\to\eta e^{in(\varphi+\varphi_0)}$, which will preserve 
the total angular momentum. In the open 
string metric for this supertube $M$ is not zero even if the tube is at the 
surface of light, but by taking $m$ and $n$ large and almost equal one can 
make $M$ arbitrarily small. If in addition we use a tube with 
$J_1=J_2$, it would have an extra modulus of rotating one of the planes 
into the other. Then, when we average over the value of this modulus, this 
supertube will smear over the entire surface of the $S^3$ domain-wall.

There is a chain of dualities \cite{Lunin:2001jy,lmm} that takes
the supertube to the D1-D5 system. If we apply this transformation
to the above metric, the exterior would be the limit of the BMPV
black hole with no linear momentum, which has eight
supersymmetries. The duality replaces the fundamental strings with
the D1-branes, so $U$ will be the relevant harmonic function,
while $V$ is the harmonic function for the D5-branes, which are
dual to the D0-branes in the IIA picture. The metric is given by
\begin{equation}
\begin{aligned}
ds^2=&U^{-1/2}V^{-1/2}\left[-(dt-A)^2 +(dy+B)^2\right]
\\&
 +U^{1/2}V^{1/2}\left[dr^2+r^2d\theta^2+r^2d\phi^2
 +r^2d\psi^2+2r^2\cos\theta\,d\phi d\psi\right]
 +U^{1/2}V^{-1/2}\sum_{i=1}^4(dx^i)^2\,,
\\
 C_2=&U^{-1}\left(dt\wedge B+dy\wedge A+2A\wedge
 B+(U-1)dt\wedge dy\right) +{\cal C}
\\
 e^{2\Phi}=&U/V\,.
\end{aligned}
\end{equation}
The field strengths $dA$ and $-dB$ are Hodge dual in the four
dimensions transverse to the branes, and the field strength
$d{\cal C}$ is dual in four dimensions to $-dV$. For our
particular choice of $A$ and $V$, we find that outside of the
domain-wall $B=-A$, and ${\cal C}=-Q_5\cos^2\theta d\phi\wedge
d\psi$. Inside the shell we get $B=A$, and ${\cal C}=0$.

This space is similar in some ways to a rotating black hole, but
it lacks a finite size horizon. Since the existence of closed time-like 
curves is a T-duality invariant notion \cite{Maoz:2003yv},
it also suffers from closed time-like curves, to
see them we should consider a vector 
$\alpha\partial_\psi+\partial_y$. This represents a closed 
curve if $y$ is compact, and we may adjust
$\alpha$ so $dy+B$ acting on this vector is arbitrarily small\footnote{
It can be set to zero, but then perhaps it would not be closed, so
one has to approximate the ratio of the period of $y$ and $2\pi/\alpha$ 
by a rational number and get a closed curve.}.
With this choice of $\alpha$, the norm squared of this vector is 
proportional to $-J^2/r^4+UVr^2$, as in the dual case. Assuming small 
$r$ this vanishes again at a radius where the angular momentum is 
related to the product of the D1 and D5 charges by $J^2=R^2Q_1Q_5$. 
Recall that if the branes form a single bound state 
the angular momentum satisfies $J=Q_1Q_5$, so the radius where 
the closed time-like curves appear is $R^2=Q_1Q_5$.

If we were not to include the
domain-wall, it would have had closed time-like curves. With the
domain-wall the interior is one of the G\"odel spaces of type IIB.

There is another possibility for the exterior metric, if we drop
the constant from the harmonic functions $U$ and $V$. This
corresponds to taking a near-horizon limit, and after the above
chain of dualities, it is the near horizon geometry of the D1-D5
system. That is $AdS_3\times S^3$ with angular momentum in
two directions \cite{Cvetic:1998xh,vijay,liat}.

\subsection{Three-charge black hole}

An even more interesting example is the three charge black hole itself
\cite{Breckenridge:1996is,Cvetic:1996xz}.
Unlike the above examples, it preserves only four supersymmetries,
not eight, and therefore the probe analysis does not apply.
Instead there would be some noncommutative gauge theory with four
supercharges, but many of the same features. We therefore will
still try to extrapolate our results to this case.

Those black holes carry three charges and two angular momenta,
we expect the singular source to be replaced, as in the previous
example by a domain-wall with the geometry $S^3\times \bR^6$.
Actually such constructions were already done in
\cite{dyson,Jarv:2002wu}. In the first, the D1 and D5-branes were
left at the origin and only a wave carrying linear and angular
momentum was put on the shell. In the second paper the D-branes
were wrapped on a K3 surface and the enhan\c{c}on mechanism was
invoked to put branes on a spherical domain-wall.

But following the example above, we expect at least the D1, the D5
and the angular momentum to sit on the shell. It seems unnatural
to leave a singular source for the momentum at the origin,
therefore it is very reasonable to assume that our mechanism
should work for a supertube-like object carrying also momentum.

Actually this object was found recently by Bena and Kraus
\cite{Bena:2004wt}. If we follow backwards the same chain of
dualities as above, it would bring the BMPV to the system of D0-D4
branes and strings with two angular momenta. Those are exactly the
charges carried by their generalized supertubes.

Again it is very simple to patch together two solutions inside and
outside of the domain-wall. In type IIB the metric of the BMPV
black hole (in ten dimensions) is
\begin{equation}
\begin{aligned}
 ds^2=&f_1^{-1/2}f_5^{-1/2}\left[-\left(dt+\frac{J}{2r^2}\sigma_L^3\right)^2
 +\left(dy+\frac{J}{2r^2}\sigma_L^3\right)^2
 +\frac{Q_k}{r^2}(dy-dt)^2\right]
\\&
 +f_1^{1/2}f_5^{1/2}\left[dr^2+r^2d\theta^2+r^2\sin^2\theta d\phi^2
 +r^2\cos^2\theta d\psi^2\right]
 +f_1^{1/2}f_5^{-1/2}dx_i^2\,,
\\
 C_2=&f_1^{-1}dt\wedge dy+f_1^{-1}\frac{J}{2r^2}\sigma_L^3\wedge(dy-dt)
 -Q_5\cos^2\theta d\phi\wedge d\psi\,,
\\
 e^{2\Phi}=&f_5^{-1}f_1\,.
\end{aligned}
\end{equation}
Here $x_i$ are the four flat directions. $f_1$ and $f_5$ are harmonic
functions
\begin{equation}
  f_1=1+\frac{Q_1}{r^2}\,,\qquad
  f_5=1+\frac{Q_5}{r^2}\,.
\end{equation}

This will be the metric outside of the domain-wall, and we place
the domain-wall at the velocity of light surface, where 
$(R^2+Q_1)(R^2+Q_5)(R^2+Q_k)=J^2$. In the over-rotating case, where
$J^2>Q_1Q_5Q_k$ this happens at positive $R^2$, while in the more realistic 
under-rotating case one has to analytically continue the metric beyond 
the horizon at $R=0$, and place the domain wall there. By putting all
the charges on the domain-wall we can find what the metric would
be inside. The result is
\begin{equation}
\begin{aligned}
 ds^2=&f_1^{-1/2}f_5^{-1/2}\left[-\left(dt+cr^2\sigma_L^3\right)^2
 +\left(dy+cr^2\sigma_L^3\right)^2
 +\frac{Q_k}{R^2}(dy-dt)^2\right]
\\&
 +f_1^{1/2}f_5^{1/2}\left[dr^2+r^2d\theta^2+r^2\sin^2\theta d\phi^2
 +r^2\cos^2\theta d\psi^2\right]
 +f_1^{1/2}f_5^{-1/2}dx_i^2\,,
\\
 C_2=&f_1^{-1}\left(dt\wedge dy+cr^2\sigma_L^3\wedge(dy-dt)\right)\,,
\\
e^{2\Phi}=&f_5^{-1}f_1\,.
\end{aligned}
\end{equation}
But now $f_1$ and $f_5$ are constants. Like in the previous
examples this is a metric that if analytically continued will have
closed time-like curves, and has no localized sources. It is a
G\"odel-like metric which also carries momentum. This is an
example of the mixed G\"odel/pp-wave metrics found in
\cite{harmark} (see for example their equation (2.50)).

One can again check the junction condition across the interface, and 
indeed in the dual type IIA metric the stress-energy tensor includes 
only terms related to the fundamental strings, D0-branes and D4-branes 
as well as an off-diagonal component related to the angular momentum 
(though in the over-rotating case the shell carries too much angular 
momentum). 
We leave for future work a fuller understanding of this geometry and 
possible relations to the metrics found in 
\cite{Lunin:2004uu,Bena:2004wv}.\footnote{These metrics were 
studied in more detail in \cite{Gimon:2004if}, which appeared right 
after the first version of this paper was posted.}

\section{Discussion}

In this paper we tried to generalize the observation in \cite{we}, 
that domain-walls made out of supertubes cutoff spaces that otherwise would 
have closed time-like curves. We demonstrated this for a few metrics 
of the G\"odel type, as well as asymptotically flat metrics with a rotating 
source. As one brings a supertube towards the velocity of light surface it 
delocalizes over the entire surface, joining the domain wall.

There has been a lot of discussion recently in the string literature on spaces 
with closed time-like curves, particularly of the G\"odel type. In all the 
examples we studied we were able to find a physical domain wall 
that changed the metric in a way that removed the closed time-like 
curves. These domain walls were also very natural from the point of 
view of the outside metric, and by accretion of supertubes one can 
create only a piece of G\"odel universe without closed time-like 
curves. Our conclusion is that the homogeneous G\"odel spaces should be 
considered unphysical, and is simply a naive analytical continuation, ignoring 
the domain wall.

It is a rather pleasing fact that string theory has a mechanism that 
prevents the appearance of certain closed time-like curves. This was 
studied here from the point of view of supertubes, which are simple probes
in many string backgrounds \cite{Gomis:2003zw}. More generally, 
the same must apply to all the dual metrics, where the supertube is 
replaced by a string, a KK monopole, or other objects. It would be very 
interesting to understand the analogous mechanism in those settings, and to see 
if they can be generalized to cases that are not dual to the supertube metrics.

This construction also gives a simple explanation of over-rotating
metrics. Following the same rules we will place a domain-wall at
the velocity of light surface, but this domain-wall would not be able to
carry all the angular momentum seen from the outside, only the
critical amount. The access angular momentum will have to come
from inside the domain-wall. But there are no sources inside the
domain-wall, just a G\"odel like space. So this extra rotation
will have to come from a singular Misner string inside this piece
of G\"odel universe. This is clearly singular and unphysical, by a
singular gauge transformation the metric can be brought back to
the critically rotating case. The secret to resolving this issue
is that we blew up the source into a sphere, while leaving only
the access rotation behind.

It should be possible to study other metrics that have closed time-like 
curves, like the black holes in G\"odel universe 
\cite{Herdeiro:2003un,Brecher:2003wq,Gimon:2003ms}. Again we 
expect a domain wall that will cure the asymptotical form of the 
metric. If the black hole is rotating, we may have to also add a domain 
wall at the inner horizon.

Those domain-walls are not the only way to eliminate closed time-like 
curves from those metrics. Instead of putting a domain-wall, it is possible 
to replace the point-like source with a localized supertube. The difference is 
that those solutions are not identical to the original metric, only asymptote 
to it very far away. Those supertube solutions were proposed as being the 
microstates of black-holes \cite{Lunin:2001jy,lmm,
Mathur:2003hj,Mathur:2004sv,Palmer:2004gu}.

If this interpretation is correct, it's not clear what the meaning of our 
domain-wall solution is. One possibility is that it is a superposition of 
different states of localized supertubes, while another is that some states 
were missed in the above counting.

We considered only spherical domain-walls, but it should be possible 
to construct less symmetric solutions, and they may be relevant for the 
entropy counting problem. In the case of the three charge black hole, there 
is a general argument \cite{Palmer:2004gu}
that localized supertubes of different shapes may not 
be enough to account for the entropy of the system. Perhaps domain-walls 
of arbitrary shapes would be important there.

\acknowledgments

I am deeply indebted to Tomeu Fiol and Joan Sim\'on for
collaboration at different stages of this project. I would not
have been able to complete it without their help. I would also
like to thank Ofer Aharony, Vijay Balasubramanian, 
Micha Berkooz, Alex Buchel, Dan Brace,
Gary Gibbons, Jaume Gomis, Troels Harmark, 
Liat Maoz, David Mateos, Rob Myers, 
Niels Obers and Federica Vian for helpful discussions. Part of this work was
done while at the Weizmann Institute, at the Benasque workshop,
the Perimeter Institute, MIT and UPenn, and I would like to thank
all those institutions for their hospitality and support.


\end{document}